\documentclass[11pt,a4paper]{article}
\pdfoutput=1
\usepackage{jcappub}
\usepackage{ifthen}
\usepackage{color}

\usepackage[T1]{fontenc}

\title{On the significance of power asymmetries in Planck CMB data at all scales\footnote{Based on observations obtained with Planck (\url{http://www.esa.int/Planck}), an ESA science mission with instruments and contributions directly funded by ESA Member States, NASA, and Canada.}}

\def\barcel{Departament de F\'isica Fondamental i Institut de Ci\'encies del Cosmos, Universitat de Barcelona, Mart\'i i Franqu\'es 1, E-08028 Barcelona, Spain}
\def\ferrar{Dip.~di Fisica, Universit\`a di Ferrara and INFN Sez.~di Ferrara, Via Saragat 1, I-44100 Ferrara, Italy}
\def\riodej{Instituto de F\'\i sica, Universidade Federal do Rio de Janeiro, 21941-972, Rio de Janeiro, RJ, Brazil}

\author[a]{Miguel Quartin}
\author[b,c]{and Alessio Notari}

\affiliation[a]{\riodej}
\affiliation[b]{\barcel}
\affiliation[c]{\ferrar}

\abstract{
We perform an analysis of the CMB temperature data taken by the Planck satellite investigating if there is any significant deviation from cosmological isotropy. We look for differences in the spectrum between two opposite hemispheres and also for dipolar modulations. We propose a new way to avoid biases due to partial-sky coverage by producing a mask symmetrized in antipodal directions, in addition to the standard smoothing procedure. We also properly take into account both Doppler and aberration effects due to our peculiar velocity and the anisotropy of the noise, since these effects induce a significant \emph{hemispherical asymmetry}. We are thus able to probe scales all the way to $\ell = 2000$. After such treatment we find no evidence for significant hemispherical anomalies along any of the analyzed directions (\emph{i.e.} deviations are less than 1.5$\sigma$ when summing over all scales). Although among the larger scales there are sometimes higher discrepancies, these are always less than 3$\sigma$. We also find results on a \emph{dipolar modulation} of the  power spectrum. Along the hemispheres aligned with the most asymmetric direction for $\,2\leq \ell \leq 2000\,$ we find a $3.3\sigma$ discrepancy when comparing to simulations. However, if we do not restrict ourselves to Planck's maximal asymmetry axis, which can only be known \emph{a posteriori}, and compare Planck data with the modulation of simulations along their respective maximal asymmetry directions, the discrepancy goes down to less than 1$\sigma$ (with, again, almost 3$\sigma$ discrepancies in some low-$\ell$ modes).
We thus conclude that \emph{no significant power asymmetries} seem to be present in the full data set. Interestingly, without proper removal of Doppler and aberration effects one would find spurious anomalies at high $\ell$, between $3\sigma$ and $5\sigma$. Even when considering only $\ell<600$  we find that the boost is non-negligible and alleviates the discrepancy by roughly half-$\sigma$.
}

\keywords{CMB theory, CMB aberration, CMB anomalies}

\begin{document}
\maketitle

\section{Introduction}
The Planck satellite~\cite{Ade:2013ktc} has provided the most accurate determination of the Cosmic Microwave Background amongst the full-sky experiments. For this reason it can be used not only for parameter extraction in a given cosmological model but also as a powerful test of the overall global isotropy of the models. Several papers, already starting a decade ago with the first public data from WMAP~\cite{Bennett:2003bz}, have claimed departures from the global isotropy. For the particular case of power asymmetries, such as differences in power in two antipodal hemispheres or of a dipolar modulation of the amplitude of the angular power spectrum, these claims are often at about the 3$\sigma$ level\footnote{Throughout this paper we make use of $\sigma$-levels instead of percentage values. I.e., $68.3\% = 1\sigma$, $95.4\% = 2\sigma$, etc.; see~\eqref{eq:pvalues-to-sigma} below.} (see, e.g.~\cite{Eriksen:2003db,Hansen:2004mj,Hansen:2004vq,Eriksen:2007pc,Bernui:2008cr,Hansen:2008ym, Hoftuft:2009rq,Paci:2013gs,Ade:2013nlj,Akrami:2014eta}), especially focusing at large angular scales, corresponding to multipoles $\ell \lesssim 60$ or in some cases $\ell \lesssim 600$. Since Planck is able to look at very small angular scales, corresponding to multipoles up to $\sim2000$, it is important to assess the existence of such anomalies on the full range of scales.

In particular~\cite{Eriksen:2003db} found that in the WMAP first year data on the range of multipoles $\ell = 2 - 40$ the power spectrum is higher in one hemisphere than in the opposite one, at  $3.0\sigma$ confidence level, compared with simulations.  Later~\cite{Eriksen:2007pc}  considered the issue of hemispherical power asymmetry in the three-year WMAP data, finding that a temperature field modulated by a dipole gives a substantially better fit to the observations than the purely isotropic model: the best-fit modulation dipole axis points toward $(l,b) = (225^\circ,-27^\circ)$, with an amplitude of 0.114 and a  significance level of $2.6\sigma$. Then~\cite{Hansen:2008ym,Hoftuft:2009rq} found that the hemispherical power asymmetry  extends to much smaller scales: for the multipole range $\ell = 2 -600$, significantly more power was found in the hemisphere centered at $(l,b) = (226^\circ,17^\circ)$ than in the opposite hemisphere. A model with an asymmetric distribution of power for $\ell = 2-600$ was found to be preferred over the isotropic model at the $2.9\sigma$ significance level. Interestingly however the best fit amplitude of the asymmetry was found to be decreasing at high $\ell$.

Strikingly in the recent Planck results it was then suggested in~\cite{Ade:2013nlj} (pre-print v1)  that the power asymmetry persists in an even larger multipole range, $\ell = 2-1500$, in the direction $(l, b) = (224^\circ, 0^\circ)$. The authors calculated the localised power in discs and introduced a measure for the asymmetry which is based on the clustering of the directions of maximal asymmetry in different multipole bins, finding a significant asymmetry at $2.7\sigma$. However this claim was questioned in~\cite{Notari:2013iva} where we have shown in simulations that the analysis has to be performed in the rest frame of the CMB, since Doppler and aberration effects are giving an important contribution to hemispherical asymmetries (see also~\cite{Burles:2006xf}). Such claim has been questioned also by~\cite{Flender:2013jja} which performs an analysis similar to the one of the present paper, although with some important differences that we will stress. Finally in~\cite{Ade:2013nlj} (pre-print v3, published) the Planck collaboration has reduced the claims of the presence of significant anomalies to the range $\ell\lesssim 600$. Given the different treatments in such papers and given the importance of the issue whether the Universe is violating or not the global isotropy, it is relevant to assess in an accurate way the presence of  anomalies using the method of estimating the power spectrum on opposite hemispheres, extending our previous analysis~\cite{Notari:2013iva} from simulations to the real data.

In this paper we build on the methods developed in~\cite{Notari:2013iva} to analyze real Planck data.  We focus on the estimation of the power spectrum on opposite hemispheres, with respect to some special directions.
In particular we analyze three different directions: the galactic North--South direction, the direction of the CMB dipole and the direction $(l,b)=(225^\circ, 1^\circ)$, which has been claimed by~\cite{Ade:2013nlj} to be the direction of maximal asymmetry, albeit obtained with different methods. Curiously, this last direction is also not far from the one that was recently found to maximize a different kind of hemispherical asymmetry, relating not the amplitude of power but to a variance dipole~\cite{Bernui:2014gla}.

Several ingredients are important in order to perform an unbiased analysis of the real Planck data of the CMB sky in addition to the boost factor analyzed in simulations in our previous work~\cite{Notari:2013iva}. First we need to deal with the fact that a mask is applied to the data, which covers the foreground signals (galaxy and point sources). This is usually taken into account of by using a mode coupling matrix $M_{\ell \ell'}$~\cite{Hivon:2001jp} (the so-called MASTER approach) which multiplies the power spectrum of the masked sky ($\tilde{C}_\ell$, the pseudo power spectrum) and reconstructs the best estimator for the original full-sky spectrum ($C_\ell$), and this is done separately for the northern and southern hemispheres. This procedure has the advantage of reducing correlations present in the pseudo power spectrum.\footnote{Such a procedure has been partially applied by~\cite{Flender:2013jja} by applying a correction for these {\it inter-scale correlations}.}  However such procedure contains a number of approximations and in order to have a cleaner procedure which is unbiased by construction we introduce a new and more conservative mask, obtained by symmetrizing the original Planck one by removing all pixels which are antipodal to the ones which are originally ignored. In this way we lose some extra fraction of the data, but we have a more reliable and clean procedure to estimate differences in opposite hemispheres, since by construction the new mask cannot introduce any asymmetry. We also take care of the fact that when cutting the sky into hemispheres this would introduce  a sharp edge, by introducing a smoothing (or apodization) of the mask on a $10'$ angular scale. After applying such mask we analyze the $\widetilde{C}_\ell$ and apply the matrix $M_{\ell \ell'}$~\cite{Hivon:2001jp}  which performs the estimate of the full $C_{\ell}$'s.

Crucially then we take into account of the Doppler and aberration effects, using a modified version of the HEALPix\footnote{\url{http://healpix.sourceforge.net/}} package~\cite{Gorski:2004by} that allows inclusion of these two effects. We assume that our velocity $\mathbf{v}$ compared with the CMB rest frame is given by the CMB dipole with
\begin{equation}\label{eq:beta}
    \beta\equiv |\mathbf{v}|/c=(1.231 \pm 0.003) \times 10^{-3}
\end{equation}
in the galactic coordinates
\begin{equation}\label{eq:betadir}
    (l,b)=(264^\circ,48^\circ)\,.
\end{equation}
Such value and direction is obtained by combining the measured temperatures of the WMAP dipole~\cite{Hinshaw:2008kr} with the COBE monopole~\cite{Lineweaver:1996xa,Mather:1998gm} and it has been also confirmed recently by the new method of estimating non diagonal correlations in the CMB~\cite{Amendola:2010ty,Notari:2011sb}. Both effects distort the CMB at all scales and induce a very significant north-south asymmetry~\cite{Burles:2006xf, Amendola:2010ty,Pereira:2010dn}. This was partially taken into account in~\cite{Flender:2013jja}, but they only subtracted Doppler without taking into account of aberration, which actually is the dominant effect at large $\ell$.

Another very important ingredient that we include in our analysis is the experimental noise, which is not symmetric (see~\cite{Ade:2013ktc}) and which is important at very small angular scales (multipoles $\ell\gtrsim 1300$). In this way we are able to perform an analysis which extends to very high $\ell$.

It is important to note that for very large scale anomalies ($\ell < 10$), other non-primordial effects such as the kinetic Doppler quadrupole, the kinetic Sunyaev-Zel'dovich effect and the integrated Sachs-Wolfe effect might be relevant~\cite{Francis-2010,Rassat:2013caa,Rassat:2013gla,Rassat:2014yna}. Here, however, we do \emph{not} focus on the very large scales, and therefore neglect these effects.

The paper is organized as follows. In section~\ref{mask} we discuss the data that we use and the treatment that we apply to the masks. In section~\ref{dopplernoise} we discuss how to take into account of Doppler and aberration and of the anisotropic noise. In section~\ref{sec:hemispherical} we discuss the significance of hemispherical asymmetries and in sections~\ref{sec:modulation}--\ref{sec:modulation2} we discuss how to put a constraint on a  specific parametrization of the north-south asymmetries, usually called dipolar modulation. We draw our conclusions in~\ref{sec:conclusions}. More specific details are listed in the appendices: in appendix~\ref{sec:U73symm} we discuss the changes introduced by our symmetrized mask, in appendix~\ref{sec:symmetrizing} we show how to build such a mask and finally in appendix~\ref{sec:healpix} the details of our HEALPix simulations.

\section{Data and symmetrized mask} \label{mask}
We analyze both data and noise presented by the first release from the Planck collaboration in the SMICA map which is a CMB temperature map obtained as a multipole-dependent linear combination of several frequency maps~\cite{Ade:2013hta}. This combined map has a power spectrum which is cosmic variance limited up to $\ell \sim 1700$, after which the noise quickly becomes dominant.

In order to minimize possible north-south or dipolar biases induced by the masking procedure, we suggest a new technique which relies on the following symmetrization of the mask. Starting from a specific mask, the symmetrized mask is obtained  by applying a parity transformation $P: \boldsymbol{\hat{n}}\rightarrow -\boldsymbol{\hat{n}}\,$
on a given pixel at a direction specified by a unit vector $\boldsymbol{\hat{n}}$ and then by multiplying by the original mask. In this way, if a pixel is absent from the original mask, the corresponding antipodal pixel will also be absent from the new mask. This is a conceptually clean way of guaranteeing that there will be no parity asymmetry in the final analysis induced by the mask. In appendix~\ref{sec:symmetrizing} we discuss how such a mask can be produced in pixel-space with HEALPix.

\begin{figure}[t]
\begin{minipage}[c]{\linewidth}
    \centering
    \includegraphics[width=\textwidth]{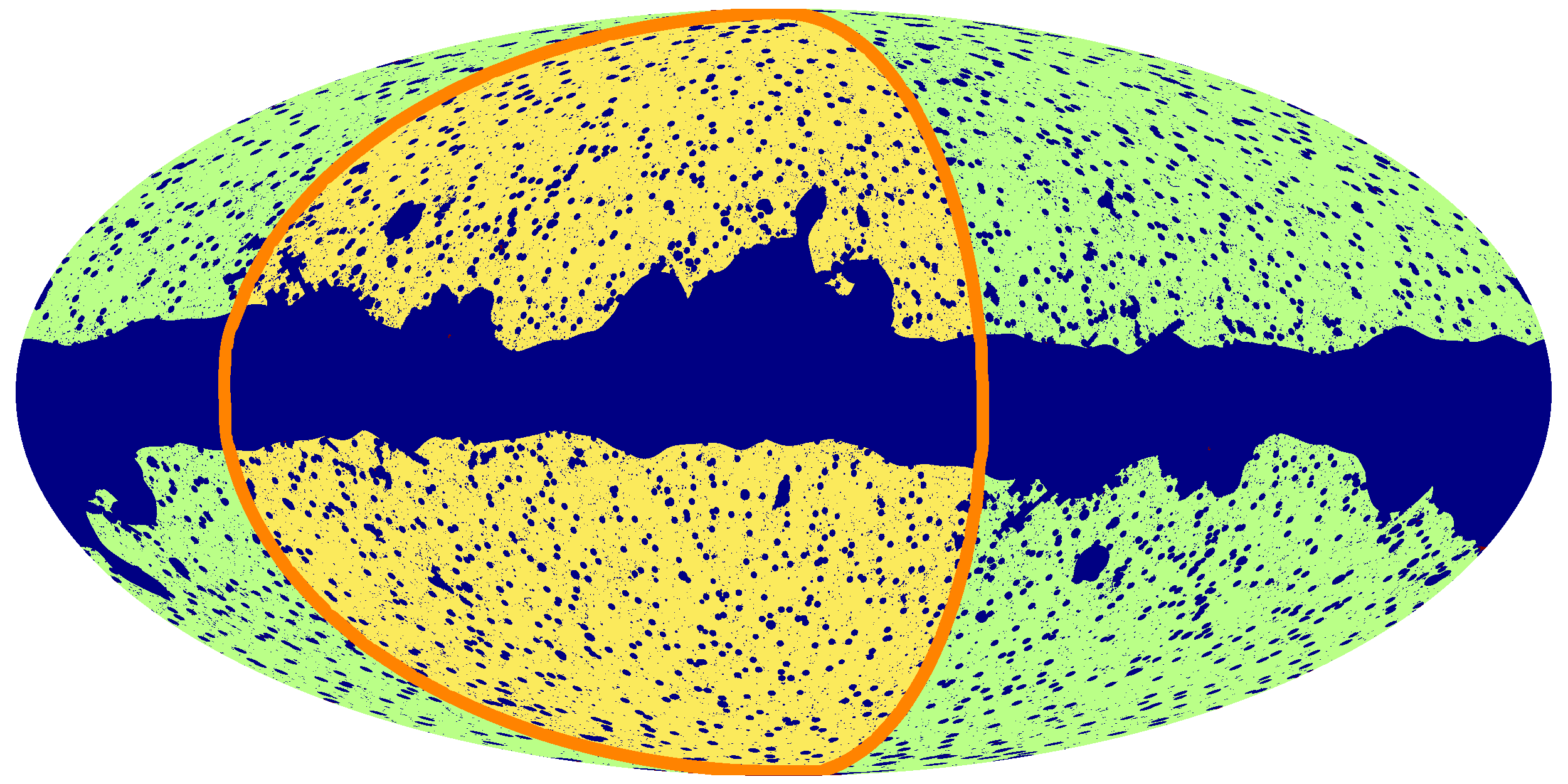}
\end{minipage}
    \caption{Antipodally-symmetric version of Planck's U73 mask. The symmetrization is such that for every masked pixel, the antipodal pixel is also masked. This removes any unaccounted bias on opposite hemispheres arising from the mask itself. It also slightly reduces the amount $f_{\rm sky}$ of unmasked sky, lowering it from 0.73 to 0.65. The green and yellow regions separated by the thick orange curve represent the two hemispheres aligned along Planck's \emph{maximal asymmetry} direction, which is one of the 3 directions analyzed in the present paper. This mask is available online [see text].}
    \label{fig:mask}
    \end{figure}

We analyze such map with the HEALPix package, with $N_{\rm side}=2048$ and $\ell_{\rm max}=4000$ (further details can be found on Appendix~\ref{sec:healpix}). We use the U73 mask, which covers the galaxy and point sources leaving $73\%$ of the sky unmasked. We dub the corresponding symmetrized mask U73symm, which is shown in Figure~\ref{fig:mask}. In other words, U73symm is the  ``antipodally symmetric'' version of U73. The downside of the symmetrization is that the leftover area of the sky is diminished, in this case to $65\%$ .

We then prepare half-sky masks, along the selected directions: North-South with respect to the galactic plane, with respect to the dipole $(l,b)=(264^\circ,48^\circ)$ and to the {\it hemispherical maximal asymmetry} direction $(l,b)= (225^\circ,1^\circ)$, originally obtained in~\cite{Ade:2013nlj}. We obtain this by rotating the U73symm with a modified HEALPix routine {\it alteralm} and cutting half of the sky with a modified version of the HEALPix routine {\it anafast} that allows for asymmetric sky cuts. These new routines are part of a modification of HEALPix which also allows the inclusion of Doppler and aberration effects, and which we now dub HEALPix-Boost.\footnote{The code has been originally developed and used in~\cite{Catena:2012hq} to check for biases in power spectrum and non-Gaussianity~\cite{Catena:2013qd}. It has also been cross-checked using analytical fitting functions which reproduce the aberration effects with high precision in~\cite{Notari:2011sb,Catena:2012hq} and further cross-checked for biases in~\cite{Notari:2013iva}.}${}^,$\footnote{The U73symm mask as well as the modified source code of HEALPix which allows for the inclusion of Doppler and aberration, are made available at~\url{www.if.ufrj.br/~mquartin/cmb}} Crucially in the final step we also smooth such half-sky masks on a $10'$ scale to avoid the presence of sharp edges, which would otherwise add significant spurious power on small scales and could either enhance or suppress artificially the presence of North-South (NS) asymmetries. For each of these masks we also obtain the MASTER matrix $M_{\ell \ell'}$, which allows to get an estimator of the full-sky $C_\ell$'s from the masked $\widetilde{C}_\ell$'s. More details on our HEALPix simulations can be found in Appendix~\ref{sec:healpix}.

Since the SMICA data and noise are beamed with a $5'$ full-width half-maximum (FWHM) Gaussian, in order to compare with primordial (unbeamed) simulations, we debeam the Planck spectrum before making comparisons. On the other hand, Planck SMICA data is deconvolved from the HEALPix ($N_{\rm side} = 2048$) pixel window function (in other words, they remove it from their data), so we take care also to not include the pixel window function in our simulation spectra.

\section{Dealing with  Doppler, aberration and anisotropy in the noise} \label{dopplernoise}

In order to check the significance of anomalies in the Planck data  we need to compare with simulations and for this purpose we performed 1000 simulations with the HEALPix routine {\it synfast}, using as a fiducial power spectrum the best-fit spectrum for a  $\Lambda$CDM model given by the
Planck collaboration~\cite{Ade:2013ktc}.

As a next step we need to take into account of the fact that the CMB is measured in a boosted frame. We perform such transformation using HEALPix-Boost, which allows introduction of boosts directly on the maps in the Planck data.  Specifically before applying the half-sky masks we  apply a boost to all simulations with velocity given by the central value of Eqs.~\eqref{eq:beta}--\eqref{eq:betadir}. Actually the boost is performed by applying the aberration with such value and by applying separately the Doppler effect multiplied by a correction factor (sometimes referred to as \emph{boost factor} $b_\nu$) of 2.5. The frequency-dependent boost factor $b_\nu$ arises from the fact that the temperature maps computed by the Planck collaboration do not exactly represent the map of the thermodynamic temperature (sometimes referred to simply as \emph{CMB temperature}). Instead, they are computed from a linear order transformation which converts intensity into temperature assuming a black-body radiation spectrum. The resulting map thus no longer transform under boosts exactly as a temperature map would, and the boost factors are the necessary corrections [up to ${\cal O}(\beta^2)$] to the Doppler effect. For the Planck temperature SMICA map, in which the signal-to-noise is dominated by the 143 GHz ($b_v \simeq 2$) and 217 GHz ($b_v \simeq 3$) channels, a simple estimation is to use the average $b_v \simeq 2.5$, as discussed in~\cite{Aghanim:2013suk}. In order to provide a detailed quantitative conclusion on the significance of the boost, half of these 1000 simulations included a boost, while the other half were traditional, unboosted simulations.

We then analyze each simulation applying 6 different masks, representing the two hemispheres along each of the 3 directions here investigated. Once this is done we extract the $\widetilde{C}_\ell^{(N,S)}$ from the North or South map and finally we also compute the $C_\ell^{(N,S)}$. An alternative procedure, which could constitute a double-check, is to deboost the masked real data (applying a boost with negative $\beta$, again employing the boost factor of 2.5) and compare the results with the simulations without any boost. The idea of deboosting the CMB sky in order to avoid systematic contamination in observables was discussed in detail in~\cite{Menzies:2004vr,Notari:2011sb}.\footnote{Ideally it should be carried out in the original data before any map-making, but this requires substantial changes to Planck's pipeline.} Since aberration changes the location of all point sources in the sky, it is important to apply any point source masks before deboosting (or equivalently to use a de-aberrated mask), otherwise foreground contamination would be incorrectly masked.  However, when deboosting the SMICA map some spurious effects are  introduced, because the noise and the edges of the masks also get deboosted, leading to some differences with the procedure of boosting the simulations. We therefore leave a full study of this possibility for a future analysis.

Finally an important issue is the one of the instrumental noise: since the observation time is not the same in all directions for Planck, there is some intrinsic anisotropy of the noise which has to be taken into account. One way of doing this is to employ the noise map for  SMICA released by Planck, which is obtained by subtracting two \emph{half-ring} SMICA maps, where such maps are simply obtained in the same way as for the SMICA full map, but considering only half of the observation time. This, however, is not optimal as it is equivalent to a single noise Monte Carlo (MC) simulation. We therefore made use instead of the 100 MC noise simulations made available from Planck. Since the Planck collaboration has not released simulations for the combined SMICA map, we made use instead of the MC simulations for the 143 and 217 GHz frequencies. These two are then combined multiplying each one by the SMICA $\ell$-dependent weight functions, following~\cite{Ade:2013hta}. As it can be seen from figure D.2 of that paper, the sum of the 143 and 217 GHz channel represent over 95\% of the total noise in the $700<\ell<2500$ range, which covers the whole range of interest here (below $\ell=1000$ noise is completely irrelevant). We then analyze the noise power spectrum in the different half-sky cuts and add the mean of the MC noise simulations to the CMB results.

To summarize, we have made 500 primordial CMB simulations containing a boost, 500 primordial CMB simulations without boost, and combined these with 100 MC noise simulations for the SMICA combined map made available from the Planck collaboration (although the Planck team produced 1000 such noise simulations, only 100 of those were made publicly available).

\section{Results: significance of Hemispherical Asymmetry} \label{sec:hemispherical}

\begin{figure}[t]
\begin{minipage}[c]{\linewidth}
    \centering
    \includegraphics[width=1.\textwidth]{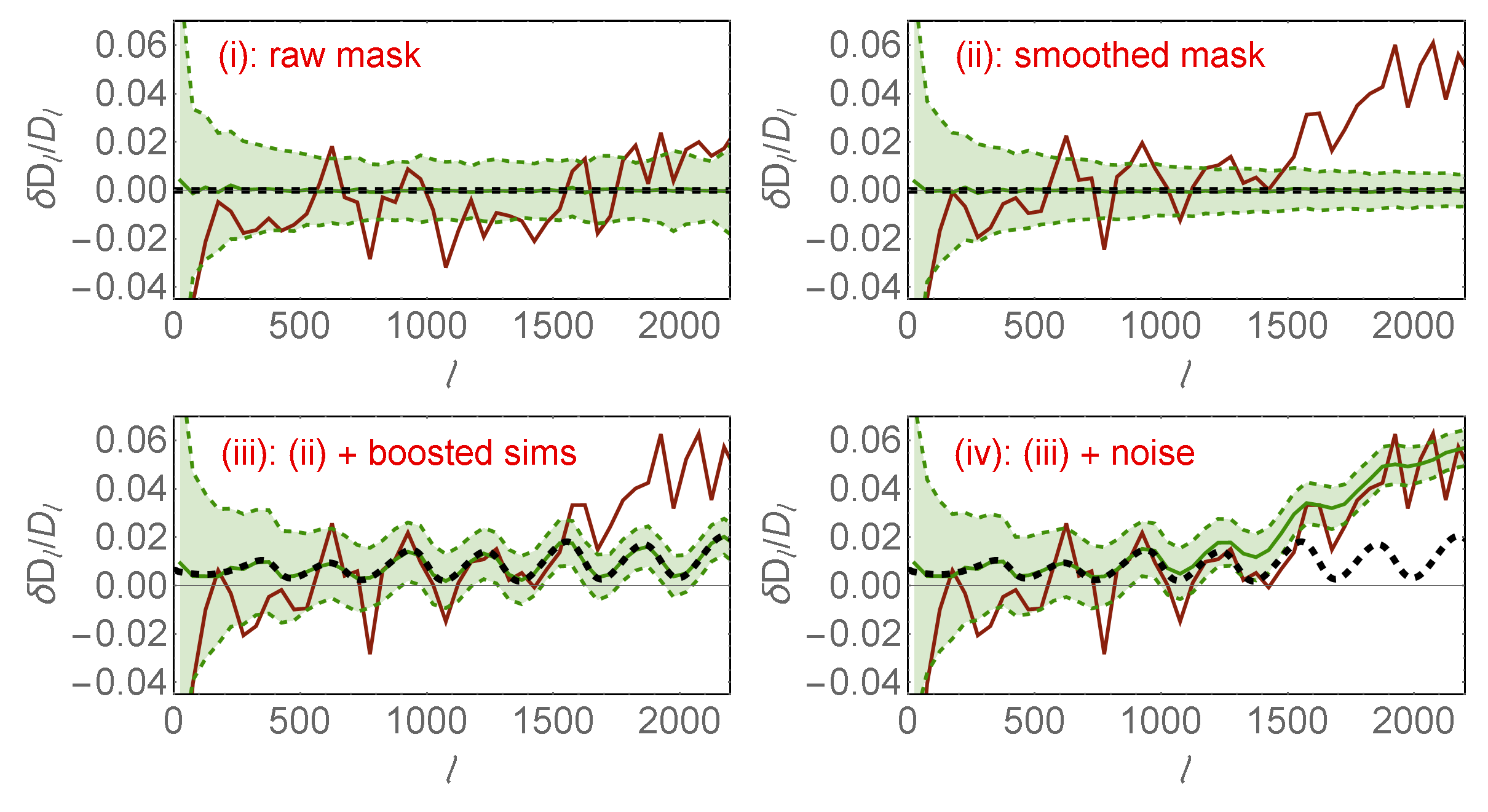}
    \end{minipage}
    \caption{The steps taken towards a correct evaluation of the power asymmetries, for the particular case of the galactic north--south hemispheres. The brown curve represents real data from Planck SMICA, the green curve and band represents the mean and $1\sigma$ bands from 500 simulations. \emph{(i):} the raw result obtained with the symmetrized U73 mask, without taking either boost or noise into consideration. \emph{(ii)}: apodizing the mask reduces the scatter in the simulations and removes biases on the real data. \emph{(iii):} the addition of Doppler shifts the simulation mean by $4\overline{\beta}$, while aberration introduce wiggles.  \emph{(iv):} the addition of noise introduces corrections for $\ell \gtrsim 1300$, introducing significant levels of anisotropy and slightly increasing the scatter.
    \label{fig:steps}}
\end{figure}

\begin{figure}[t]
\begin{minipage}[c]{\linewidth}
    \centering
    \includegraphics[width=\textwidth]{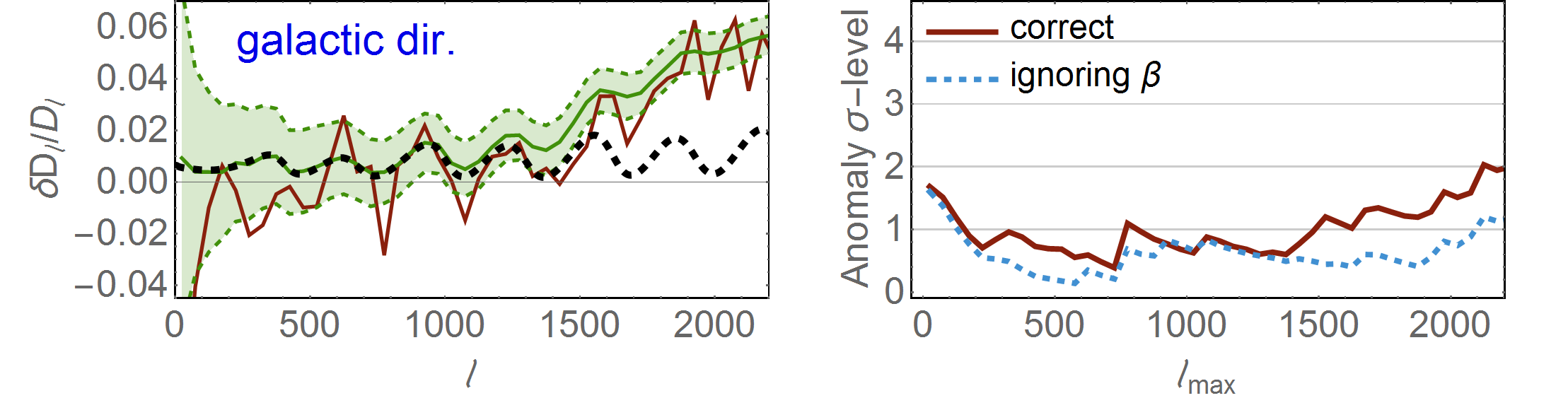}
    \includegraphics[width=\textwidth]{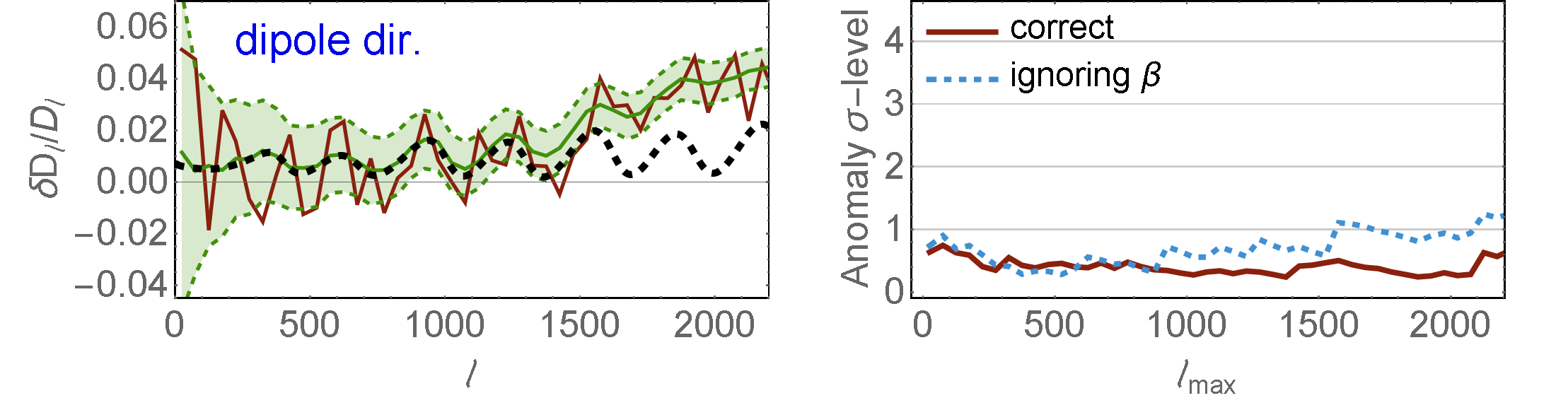}
    \includegraphics[width=\textwidth]{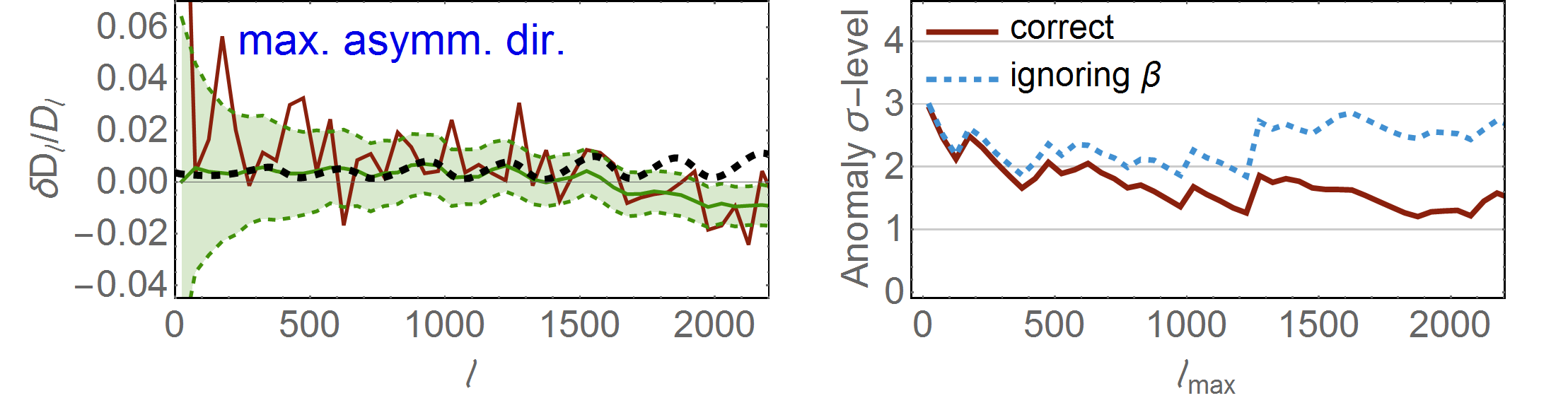}
\end{minipage}
    \caption{Hemispherical asymmetry and its statistical significance along 3 different directions. \emph{[Left:]} the relative difference between $D_{\ell}$'s in two opposite halves of the sky as a function of the multipole $\ell$.  The brown is the Planck data while the green curve and band represent the mean and $1\sigma$ band from 500 simulations, binned in 50$-\ell$ bins. The black dashed curves are the analytical estimate of Eqs.~\eqref{eq:dAoverAest}--\eqref{averagebeta}, which ignores all noise.
    \emph{[Right:]} the corresponding statistical (in)significance of the anomaly, summing all multipoles from 2 to $\ell_{\rm max}$. The brown curve represents the correct estimate; the dotted blue curve is the spurious significance if one ignores to boost the simulations.  Note that naively ignoring the boost can lead to a spurious anomaly at high $\ell$ between $2.5$ and $2.9\sigma$.
    \label{significance}}

\end{figure}

For each of the above mentioned three directions we finally compute the spectra $D^{(N,S)}_\ell \equiv C^{(N,S)}_\ell \ell (\ell+1) / (2\pi)$, and depict the difference
\begin{equation}
    \frac{\delta D_\ell}{D_\ell} \;\equiv\; \frac{D^N_\ell - D^S_\ell}{D_\ell^{\rm average}} \;\equiv\; 2 \;\frac{D^N_\ell - D^S_\ell}{D^N_\ell + D^S_\ell}
\end{equation}
in Figs.~\ref{fig:steps} and~\ref{significance}, comparing with mean and standard deviation of the same quantity for the simulations. Note that for simulations the $D_\ell$'s contain also the boost effects and the noise power in each hemisphere. All results are obtained  after binning the spectra in 50-$\ell$ bins. Specifically in Figure~\ref{fig:steps} we show step by step how important is each of the effects that we consider: smoothing of the mask, boost (Doppler and aberration) and the inclusion of the anisotropic noise. In both figure we also depict (dashed black curve) the average bias due to Doppler and Aberration, which is oscillating and has non-zero mean. Such bias was found in~\cite{Notari:2013iva} to be extremely well approximated by the analytical expression
\begin{equation}\label{eq:dAoverAest}
    \frac{\delta D_\ell}{D_\ell}  \;\simeq\;  4 \overline{\beta} \,+\, 2 \overline{\beta} \, \ell \, \bigg(1-\frac{D_{\ell+1}^{\rm th}}{D_{\ell}^{\rm th}}\bigg)\,,
\end{equation}
where  $D^{\rm th}_\ell$ is the fiducial power spectrum and $\overline{\beta}$ is the average
\begin{equation}
    \overline{\beta}=\beta \langle \cos{\gamma} \rangle_R=\beta \int_{R} d\Omega \cos{\gamma} \label{averagebeta}
\end{equation}
where in turn $R$ is the region of interest integrated over the solid angle $d\Omega$ and $\gamma$ is the angle relative to the boost direction. We can also see that the simulations at $\ell\gtrsim 1300$ start  having a net positive bias  and this is due to the anisotropic noise which has been added to them and which becomes important at such high $\ell$.

In Figure~\ref{significance} we show the final results for the three directions and include also the statistical significance of the asymmetry summing from $\ell = 2$ to different $\ell_{\rm max}$, in order to illustrate its evolution as we probe ever smaller scales. This significance in turn could in principle be computed using the fact that the $a_{\ell m}$'s are Gaussianly distributed, which implies that the $D_\ell$'s are distributed according to a $\chi^2$ distribution. However, we verified numerically that the binned quantities $\delta D_\ell/D_\ell$ themselves are very well described by Gaussian distributions. Therefore, for $i$ multipole bins we have a simple $\chi^2$ distribution with $i$ degrees of freedom:
\begin{equation}
    \chi^2_{i}\;=\; \sum_{{\rm bin}=1}^{i}\frac{\Big[ \big(\delta D_\ell/D_\ell\big)_{\rm Planck} -  \big(\delta D_\ell/D_\ell\big)_{\rm sims}  \Big]^2 }{ \sigma_{\rm bin}^2}\,,
\end{equation}
where $(\delta D_\ell/D_\ell)_{\rm sims}$ represents the average over the simulations and $ \sigma_{\rm bin}^2$ is the variance of $(\delta D_\ell/D_\ell)$ in a given bin. This variance is a sum of a term arising from cosmic variance and another from the instrumental noise. For a given $D_\ell$ the uncertainty is given by (see~\cite{Hivon:2001jp} or Eq.~(11.27) in~\cite{Dodelson:2003ft})
\begin{equation}
    \sigma_{D_\ell} \;=\; \sqrt{\frac{2}{(2\ell +1)f_{\rm sky}}} (D^{\rm th}_\ell + N_\ell^{\rm MC})\,,
    \label{sigma2}
\end{equation}
where $N_\ell^{\rm MC}$ is the noise power spectrum (obtained as an average through Monte Carlo simulations) multiplied by $\ell(\ell+1)/(2\pi)$. In principle, there are extra factors in~\eqref{sigma2} due to the mask apodization~\cite{Hivon:2001jp}, but for U73symm these are negligible. Propagating this error into the quantity $\left(\delta D_\ell / D_\ell \right)$, we get for a bin of $\Delta$ multipoles
\begin{equation}
    \sigma_{\rm bin}^2 \;=\; \frac{1}{\Delta} \frac{\big(D_\ell^N \sigma_{D_\ell}^N\big)^2 + \big(D_\ell^S\sigma_{D_\ell}^S\big)^2}{\big(D_\ell^{\rm average}\big)^4}
    \;\simeq\; \frac{1}{\Delta} \frac{4\,\big(1-\rho^{\rm NS}_\ell\big)}{(2\ell+1)f_{\rm sky}} \,,
\end{equation}
where $\rho^{\rm NS}_\ell$ is the correlation between the north and south hemisphere spectra in a given $\ell$-bin. In the last step above we used the fact that (as will be shown) the asymmetry between the two hemispheres is never beyond a few percent to approximate $D_\ell^N \simeq D_\ell^S \simeq D_\ell^{\rm average}\simeq D^{\rm th}_{\ell} + N_{\ell}^{\rm MC}$.

With the total $\chi^2$ we perform a goodness-of-fit test to see whether the isotropic standard model is a good fit to the data. Finally, to quote the significance in $\sigma$ levels (instead of p-values, or probabilities to exceed) using all bins from the first to the $i$-th bin, one must convert the probabilities given by
\begin{equation}
    {\rm prob} = {\rm CDF}\big[\chi^2, \,i\big] = 1-\textrm{p-value}\,.
\end{equation}
In what follows we refrain from quoting results in terms of p-values, preferring instead using the corresponding $\sigma$-levels, which are \emph{defined} through the Gaussian distribution:
\begin{equation}\label{eq:pvalues-to-sigma}
    \sqrt{2} \,{\rm Erf}^{-1} \big(\textrm{prob} \big)\,.
\end{equation}
Here, ${\rm Erf}^{-1}$ is the inverse of the error function and ${\rm CDF}$ is the cumulative distribution function of the $\chi^2$ distribution with $i$ degrees of freedom:
\begin{equation}
    {\rm CDF}[x, \,i] = \frac{\gamma[i/2,x/2]}{\Gamma[i/2]}\,,
\end{equation}
where $\Gamma$ and $\gamma$ are the gamma the incomplete gamma function, respectively. In other words, this is a straightforward generalization of the commonplace association of $1,\,2,\,3\sigma$ as shorthand notation for $68.3\%,\,95.4\%,\,99.73\%,$ (independently of whether the underlying distribution is really Gaussian) and so forth.

As it can be seen from the plots it is crucial to add the anisotropy of the noise and, to a smaller extent, the boost effects. Taking into account of these two sources of bias makes the real data compatible with the simulations and in fact there is no highly significant asymmetry left in any of the three directions. Note that since for $\ell<900$ the boost bias is smaller than the uncertainty, the goodness-of-fit results are never changed by more than $1\sigma$ when boost effects are included. Nevertheless, a proper account of the boost does reduce the discrepancy of the maximal asymmetry direction for $\ell \in [2,\,2000]$ from $2.5\sigma$ to $1.3\sigma$, a very important result. This possibility was already stressed in~\cite{Notari:2013iva} on simulations. Note that the boost can also increase the discrepancy, and in fact that is what happens along the galaxy direction, where we see a rise in the discrepancy in $\ell \in [2,\,2000]$ from $0.9\sigma$ to $1.6\sigma$. This is, however, still insignificant. So accounting for all directions, the proper accounting of boost does reduce the worst case discrepancy to less than $2\sigma$, also non-significant.

The only exception to the above claim can be seen in the very first bin ($2\le \ell \le 50$) along the maximal asymmetry direction. There we see a discrepancy of $3.0\sigma$. This is in very good agreement with earlier results, dating back over a decade~\cite{Eriksen:2003db,Hansen:2004mj,Hansen:2004vq,Eriksen:2007pc,Bernui:2008cr,Hansen:2008ym, Hoftuft:2009rq,Paci:2013gs,Ade:2013nlj,Akrami:2014eta}. Note also that the behaviour of the significance of Figure~\ref{significance} is in qualitative agreement with the significance of anomalies found by the Planck collaboration using a very different method (clustering of preferred directions), as it can be seen comparing with fig.28 of  v3 of~\cite{Ade:2013nlj}.

\section{Results: Dipolar modulation -- a posteriori direction} \label{sec:modulation}

Several previous papers have analyzed a parameterization, in which the CMB temperature is assumed to be modulated by a dipolar term of the form:
\begin{equation}
    T \,=\, T_{\rm isotropic} \big(1+A_{\rm mod} \cos{\alpha}\big) \,,
\end{equation}
where $\alpha$ is a relative angle between a specific direction and the direction of observation, which then translates on the $C_{\ell}$'s to lowest order:
\begin{equation}
    C_{\ell} \,=\, C_{\ell, \, \rm isotropic} \big(1+2A_{\rm mod} \cos{\alpha}\big) \,. \label{A-Cell}
\end{equation}
Note that such a parametrization corresponds exactly to the physical Doppler effect due to a boost (but does not contain aberration), where $A_{\rm mod}$ is equal to the velocity $\beta$. Although such an ansatz  is arbitrary we still provide results on the significance of a nonzero $A_{\rm mod}$, because several analyses~\cite{Hansen:2008ym,Eriksen:2007pc,Hoftuft:2009rq,Ade:2013nlj} have claimed a detection of a large value for $A_{\rm mod}$, about $60$ times bigger than the Doppler term and with a significance of about 3$\sigma$ when analyzing the low-$\ell$'s of the CMB, namely up to $\ell\lesssim 64$ or $\ell \lesssim 600$. Of course, if we consider the above parametrization there is no reason why we should stop at low $\ell$'s and the full range of $\ell$ should be used. We repeat this for the two different directions, namely  the dipole direction and the \emph{maximal asymmetry} direction. Note however that the latter is a specific direction chosen \emph{a posteriori}, since it has been already found to be the maximal asymmetry direction in Planck by other analyses. Therefore the real significance will become lower if considering the fact that the maximal asymmetry direction should be searched for in each simulation and marginalized over. In this section we compute the naive significance level, while in the next section we compute the unbiased significance level by looking for the modulation along the axis of maximal asymmetry for each simulation.

It is important to note that the data contain a nonzero value for $A_{\rm mod}$  due to Doppler, aberration and to the anisotropic noise. Such effects are contained in our simulations, therefore what we need to constrain is the quantity
\begin{equation}
    A_{\rm mod} \equiv A^{\rm Planck}_{\rm mod} - A^{\rm sims}_{\rm mod}\,.
\end{equation}
Given a half-sky cut a modulation such as~\eqref{A-Cell} will induce an average effect on the different $C^{N,S}_\ell$ given by
\begin{equation}
    C^{N,S}_{\ell}\,=\,C^{N,S}_{\ell, \, \rm isotropic} \big(1\pm 2A_{\rm mod} \langle \cos{\alpha} \rangle \big) \,,
\end{equation}
where the $\langle ... \rangle$ is the average over one hemisphere of the modulus of the cosine of the angle between a given direction and the preferred direction. Then we may introduce a $\chi^2$ as follows:
\begin{equation}
    \chi^2(A_{\rm mod}
    ) \,=\, \sum_\ell \sum_{i,j}\,\mathcal{I}^{ij}_\ell \Big( D^{i}_\ell-\overline{D}_\ell^i\Big) \Big( D^{j}_\ell-\overline{D}^j_\ell\Big)\,,
\end{equation}
where the $\mathcal{I}$ is the inverse of the covariance matrix and the indices $i,j$ can be either $N$ or $S$. Then,  $D_\ell^i$ are the quantities taken from the Planck real data or from a full simulation, while the $\overline{D}_\ell^i$ are the averages
\begin{equation}
    \overline{D}^{N,S}_\ell\,=\,\big(\overline{D}^{\rm gauss}_\ell \pm \delta D^{\beta}_\ell \big) (1\pm  2 \langle \cos{ \alpha} \rangle A_{\rm mod})+N^{N,S}_\ell \,,
\end{equation}
with $\overline{D}^{\rm gauss}_\ell $ standing for the mean of a Gaussian isotropic simulations
and where $\delta D^\beta_\ell$ is the mean effect of boost on the $N$ hemisphere which is given by~\eqref{eq:dAoverAest}:
\begin{equation}
    \delta D^\beta_\ell  \;\simeq\;  2 \overline{\beta} \overline{D}^{\rm gauss}_{\ell}\,+\, \overline{\beta} \, \ell \, \bigg(\overline{D}^{\rm gauss}_\ell-\overline{D}^{\rm gauss}_{\ell+1}\bigg)\,.
\end{equation}
In the absence of $NS$ correlations, the quantity $\mathcal{I}^{NN}_\ell=\mathcal{I}^{SS}_\ell$ would just be $1/\sigma_{D_\ell}^2$, given by~\eqref{sigma2}. The correlations offset these values, but not by much, and we have checked that neglecting them leads to only slightly different results. The estimator for the best fit value of $A_{\rm mod}$ is obtained by setting $\partial \chi^2/ \partial A_{\rm mod}=0$, which leads to the estimator:
\begin{equation}
    \begin{aligned}
    \bar{A}_{\rm mod} \,\equiv\,
    \sum_\ell \Bigg[& \overline{D}^{\rm gauss}_\ell \Big[  \big(\mathcal{I}^{NN}-\mathcal{I}^{NS}\big) \big(D^{\rm N}_\ell-D^{\rm S}_\ell-N^N_\ell+N^S_\ell \big) -4\, \mathcal{I}^{NN} \delta D^\beta_\ell \Big] \\
    &+ \delta D^\beta_\ell \big(\mathcal{I}^{NN}+\mathcal{I}^{NS}\big) \big(D^{\rm N}_\ell+D^{\rm S}_\ell-N^N_\ell-N^S_\ell \big) \Bigg] \\
    &\!\!\!\!\!\!\!\!\!\!\!\!\!\!\times \Bigg[
    4 \langle \cos{ \alpha} \rangle \sum_{\ell}  \bigg[\big(\mathcal{I}^{NN}-\mathcal{I}^{NS}\big) (\overline{D}^{\rm gauss}_\ell)^2 + \big(\mathcal{I}^{NN}+\mathcal{I}^{NS}\big) (\delta D^\beta_\ell)^2 \bigg] \Bigg]^{-1} \,,
    \end{aligned}
    \label{eq:biasAest}
\end{equation}
where the sum is performed over all the multipole bins of interest. In the absence of noise, boost and correlations between the hemispheres, this estimator would simplify to the following~\cite{Notari:2013iva}
\begin{equation}
    \bar{A}_{\rm mod} \,=\,
    \frac{1}{4 \langle \cos{ \alpha} \rangle} \frac{ \sum_\ell (2\ell+1) \delta D_{\ell}/\overline{D}_{\ell} }{    \sum_{\ell} \left(2\ell+1  \right)}\,.
    \label{eq:biasAest-nonoise}
\end{equation}
Note that this differs from the estimator used in~\cite{Flender:2013jja} because of the factors $(2 \ell+1)$ included in the sum, which weigh for the number of $m-$modes at each $\ell$.

\begin{figure}[t!]
\begin{minipage}[c]{\textwidth}
    $\!\!\!\!$\includegraphics[width=1.05\textwidth]{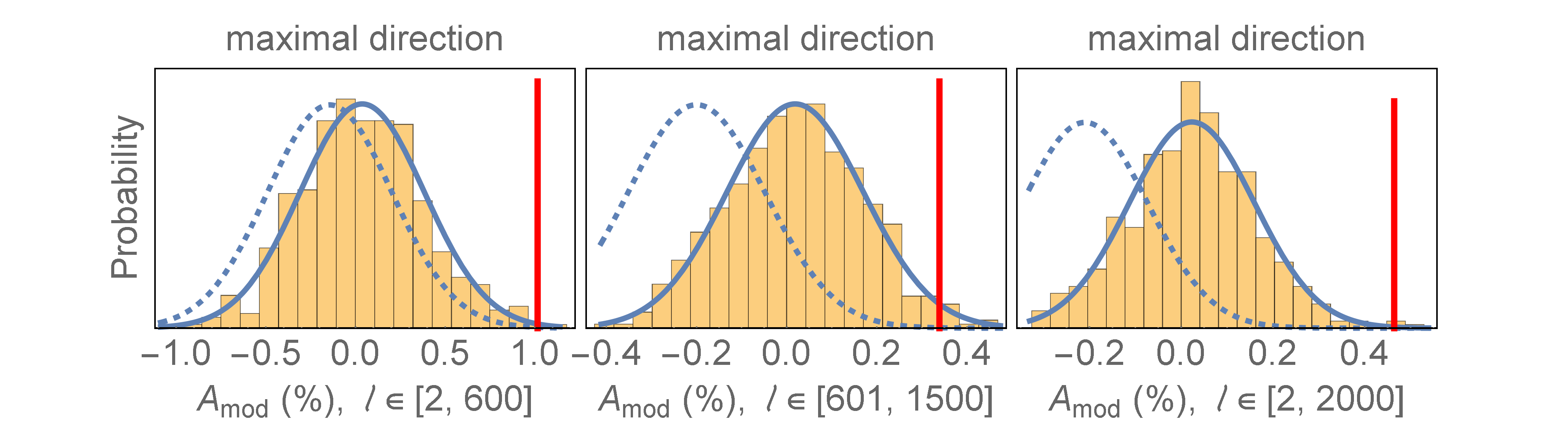}

    $\!\!\!\!$\includegraphics[width=1.05\textwidth]{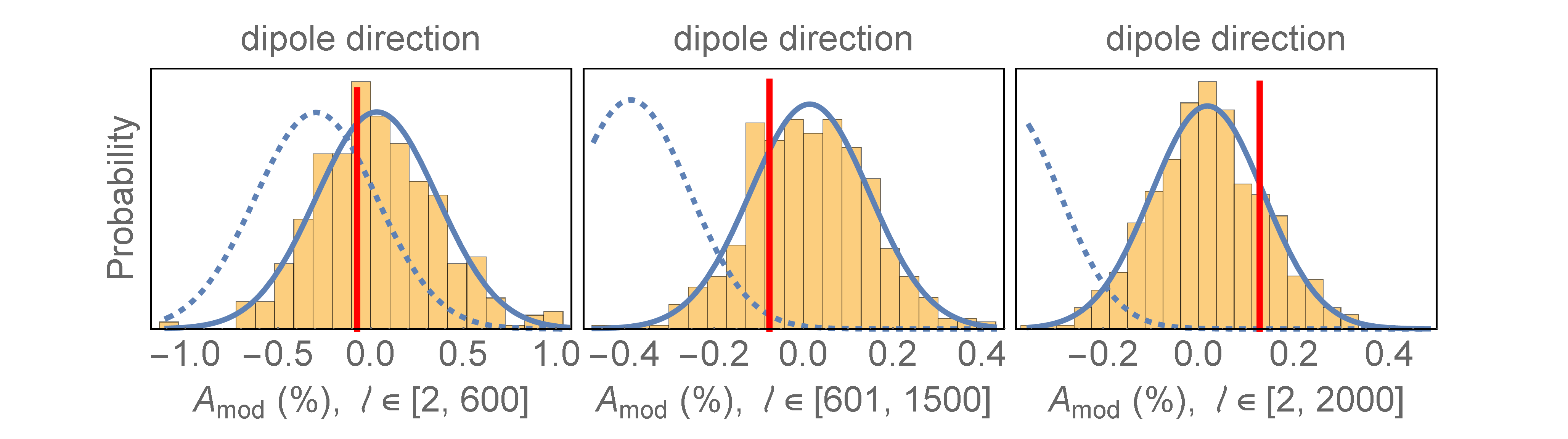}
 \end{minipage}
        \caption{Measurement and simulations of the amplitude $A_{\rm mod}$ of a possible dipolar modulation of the CMB temperature power spectrum for different multipole ranges and assuming 2 different fiducial modulation directions:  (i) the one of maximal asymmetry [see text]; (ii) the dipole one. The red vertical line is the value obtained from the Planck SMICA data; the histograms represent 500 different simulations and are in turn well-fitted by a Gaussian (solid blue line). The dashed blue line is the best-fit Gaussian for the simulations that neglect the boost. As it can clearly be seen, neglecting the boost makes the data look anomalous at high-$\ell$. Note that even for $\ell \in [2, 600]$ (roughly the WMAP range), the boost is relevant and alleviates the discrepancy in the maximal direction by $0.4\sigma$.  Note that the values along the {\it maximal asymmetry} direction contain an {\it a posteriori bias}, since the direction has been chosen precisely because it maximizes the asymmetry, and therefore the real significance becomes lower if considering the fact that the maximal asymmetry direction should be searched for in each simulation. Such an unbiased analysis in depicted in Figure~\ref{fig:amod-vs-ell-betacart}. \label{dipolar}}
\end{figure}

\begin{figure}[t]
    $\!\!\!\!\!\!\!\!\!$\includegraphics[width=1.09\textwidth]{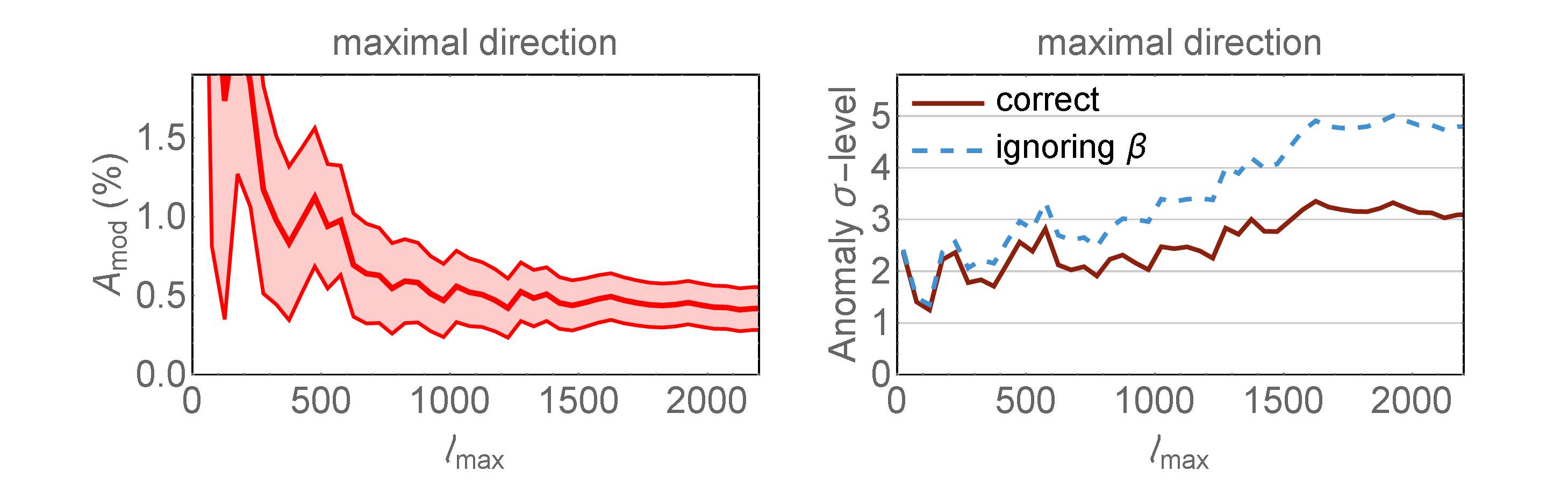}
    \caption{Similar to Figure~\ref{significance} for the measurement of $A_{\rm mod}$ in the range $\ell\in[2,\ell_{\rm max}]$. Here the discrepancy is larger and amounts to $3\sigma$ when considering all scales (and naively $5\sigma$ if one neglects boost effects). Note also that data is well fit by a constant $A_{\rm mod}$ (the very first bin $\ell \in [2,50]$ however disagrees with a constant $A_{\rm mod}$ at $2.3\sigma$).\label{fig:amod-vs-ell}}
\end{figure}

The results on $A_{\rm mod}$ are shown in Figures~\ref{dipolar} and in Table~\ref{tab:dipolar-mod} for different multipole ranges and assuming two different fiducial modulation directions:  (i) $(l,b)=(225^\circ, 1^\circ)$, found in~\cite{Ade:2013nlj} to maximize the hemispherical asymmetry; (ii) the dipole direction. Note that $(l,b)=(225^\circ, 1^\circ)$ is also very close (actually within the error bars) to the direction $(l,b) = (226^\circ,-17^\circ)$, found in~\cite{Ade:2013nlj} to maximize this dipolar modulation. We split the data intro 3 ranges: $\ell\in[2,600]$, which corresponds to the cosmic-variance-limited region of WMAP, $\ell\in[601,1500]$ which are the extra modes precisely measured by Planck (well within the cosmic-variance region) and $\ell\in[2,2000]$, which is the full range of modes accessible ($\ell \gtrsim 2000$ are clearly dominated by noise). Figure~\ref{fig:amod-vs-ell} explores further the maximal asymmetry direction by depicting the measured $A_{\rm mod}$ using $\ell \in [2, \ell_{\rm max}]$ as a function of $\ell_{\rm max}$, together with the corresponding discrepancy with the fiducial, isotropic model. As it can clearly be seen in both figures, neglecting the boost makes the data look anomalous at high-$\ell$. When properly treating the boost   we do not find any significant detection for a modulation along the dipole direction, but there is still some nonzero value along the \emph{maximal asymmetry} direction, where for the full range of scales we find $A_{\rm mod} = 0.0044\pm 0.0014$. Such result has an error and a central value which is one order of magnitude smaller than the official Planck results~\cite{Ade:2013nlj} which is limited to scales of about $\ell\lesssim 100$, and deviates from zero at the 3.2$\sigma$ level.

The 2 smaller subsets of $\ell$'s also allow comparison with previous results in the literature. For  $2<\ell<600$ it can be seen from Table~\ref{tab:dipolar-mod} that along the maximal asymmetry direction on these largest scales  we find a $2.9\sigma$ deviation from zero, with a  central value of about $A_{\rm mod} \approx 0.01$ along the \emph{maximal asymmetry} direction, which is roughly consistent with previous results~\cite{Hansen:2008ym,Eriksen:2007pc,Hoftuft:2009rq,Ade:2013nlj,Flender:2013jja,Axelsson:2013mva}.
Contrary to what is claimed in~\cite{Axelsson:2013mva}, however, we find that even on these large scales the boost effects cannot be completely neglected, and account for roughly $0.4\sigma$ of the total discrepancy. For $601<\ell<1500$ we can compare directly with~\cite{Flender:2013jja}, where a strong \emph{scale-dependent} modulation was invoked. In fact~\cite{Flender:2013jja} finds that for $\ell\lesssim 600$ there is a 3$\sigma$ anomaly, while for $601\leq\ell\leq 2048$ there is only a $1\sigma$ deviation, so  they claimed data contained a strong scale dependent modulation. We make here  two crucial objections to such statement. First, as we stressed out, our treatment includes properly many physical effects which are important especially at high $\ell$: Doppler, aberration and anisotropic noise, in addition to using a  symmetrized mask  and the MASTER matrix for reconstructing the real $C_\ell$'s.
This  more careful treatment yields a lingering $2.2 \sigma$ deviation on $600\leq\ell\leq 1500$ ($A_{\rm mod} = 0.0032\pm 0.0015$). Second, there is no reason why the data should be split into these two arbitrary subsets to constrain the single parameter $A_{\rm mod}$. In fact from Figure~\ref{fig:amod-vs-ell} we can conclude using the all data up to 2000 that there is no clear sign of a strong scale dependence for $\ell\gtrsim 300$. Indeed, the very first bin is the only one in tension, at $2.3\sigma$, in agreement with what was found in~\cite{Ade:2013nlj}. However, since for larger scales the variance increases fast, a constant $A_{\rm mod}$ is not a bad fit when considering all scales together. So contrary to~\cite{Ade:2013nlj} we find here no strong reason to claim that a possible modulation must be scale-dependent.
However note that in the next section, as discussed below, we also find that the significance goes drastically down when including $\ell>600$.

\begin{table}[t]
    \centering
    \begin{tabular}{c|c|ccc}
      \hline \hline
      \textsc{direction} & \textsc{sims} & $\ell \in [2,600]$ & $\ell \in [601,1500]$ & $\ell \in [2,2000]$ \\ \hline
       maximal& with $\beta$ & $0.98\pm0.35$ & $0.32\pm0.15$ & $0.44\pm0.14$ \\
       asymmetry 
        &  & $(2.9\sigma)$ & $(2.2\sigma)$ & $(3.3\sigma)$\\ \cline{2-5}
       &ignoring $\beta$ & $1.16\pm0.35$ & $0.54\pm0.15$ & $0.67\pm0.14$ \\
       & &  $(3.3\sigma)$ & $(3.5\sigma)$ & $(4.9\sigma)$ \\\cline{2-5}
       \hline\hline
       dipole& with $\beta$& $-0.11\pm0.33$ & $-0.09\pm0.13$ & $0.11\pm0.12$ \\
       & & $(0.2\sigma)$ & $(0.6\sigma)$ & $(1.1\sigma)$\\ \cline{2-5}
       &ignoring $\beta$ & $0.22\pm0.33$ & $0.31\pm0.13$ & $0.54\pm0.12$ \\
       & & $(0.7\sigma)$ & $(2.4\sigma)$ & $(4.6\sigma)$ \\\cline{2-5}
       \hline\hline
    \end{tabular}
    \medskip
    \caption{Constraints on $100 A_{\rm mod}$ for different multipole ranges along 2 different fiducial directions: the one found in~\cite{Ade:2013nlj} to maximize the low-$\ell$ modulation and the direction of the dipole. For each direction the first row is the most accurate estimate of $A_{\rm mod}$ because it takes into account the effect of the boost in the simulations. However note that the values along the {\it maximal asymmetry} direction contain an {\it a posteriori} bias, since it has been chosen precisely because it maximizes the asymmetry, so numbers in the first row should be regarded as upper bounds to the total asymmetry [see text].    }
    \label{tab:dipolar-mod}
\end{table}

It is in fact very important to note the following facts. First, as stressed before, the values along the maximal asymmetry direction contain an \emph{a posteriori} bias, since the direction has been chosen precisely because it maximizes the asymmetry, and therefore the real significance will become lower when considering the fact that the maximal asymmetry direction should be also searched for in each simulation (and marginalized over) to have a statistically meaningful result. We perform such an analysis in the next section, by using a different estimator, which finds the modulation along the maximal asymmetry axis.

Second, note also that a detection of non-zero $A_{\rm mod}$ does not necessarily mean that the model with dipolar modulation fits better the data, because it has one extra  parameter. In order to assess better this we conduct a Bayesian model comparison by computing the Bayes Factor $B$, which is the ratio of Bayesian Evidences (see~\cite{Trotta:2008qt}) of the models with and without $A_{\rm mod}$:
\begin{equation}\label{eq:bayesfactor}
    B \;=\; \frac{{\displaystyle\int}\exp\Big( -\frac{1}{2}\chi_{\rm A_{mod}}^2  \Big) \;{\rm prior}(A)\;{\rm d}A\,}{\exp\left( - \frac{1}{2}\chi_{\rm no \;A_{mod}}^2 \right)} \,,
\end{equation}
where in the modulated model we assumed a flat prior for $A_{\rm mod}$. Since the evidence is proportional to the probability that we associate to the correctness of the model, if we in turn assume these 2 models constitute the complete set of possibilities, then $p(\mathcal{M}_{A}) + p(\mathcal{M}_{0}) = 1$, which together with~\eqref{eq:bayesfactor} allows us to compute both probabilities~\cite{Trotta:2008qt}. We can then ask by how much is the modulation model favored. It is important to note that while the exact range of any sufficiently broad priors do not affect any Bayesian parameter estimation, the same is not correct when doing model selection analysis, such as with the Bayes Factor. In fact, other things being equal, models with broader priors are disfavored because they are less predictive. This is because of the Occam's Razor factor, which is embedded in the Bayes Factor (see~\cite{Gregory:2010} for a more in-depth discussion). Since we do not want to test specific models for a dipolar modulation here, we will assume as prior for $A_{\rm mod}$ the range $[ -1, 1 ]$. We thus get for $\ell \in [2,\,2000]$ that the modulation model, fixing the direction to be the maximal asymmetry one, is favored with $B = 54$, which corresponds to a p-value of 0.18 or equivalently, using~\eqref{eq:pvalues-to-sigma}, to a $2.4\sigma$ preference for a modulation.\footnote{In Bayesian model comparison the Jeffreys' scale is often employed, but we find it much more intuitive to discuss results in terms of $\sigma$-levels, as one of us first proposed in~\cite{Castro:2014oja}.} As expected, this is less than the naive expectation from Table~\ref{tab:dipolar-mod} ($3.3\sigma$) because the Bayes factor takes into account the fact that the modulation model has one free parameter for which we have little information a priori.

Finally note that on the $\ell<600$ range, which roughly corresponds also to the WMAP range, the boost is already non-negligible and alleviates the discrepancy from $3.3\sigma$ to $2.9\sigma$, with a central value of $A_{\rm mod}$ shifting down by $\approx 0.002$. These numbers, however, cannot be directly applied to WMAP data because there the average boost factor differs from the one in Planck.

\section{Results: Dipolar modulation -- unbiased direction} \label{sec:modulation2}

In the previous section we have assessed the significance of an anomalous dipolar power modulation of the temperature spectrum along determined directions. Actually however the direction of maximal asymmetry should not have been pre-determined (\emph{i.e.}, before looking at the data), and as such constitutes an {\it a posteriori} statistic. In order to circumvent this, one would have to repeat the analysis over hundreds of different directions for each simulation, finding its maximal asymmetry direction. This is, however, a computationally intensive route.

We thus proceed along a different path which automatically finds the modulation along the maximal asymmetry direction, by making the crucial observation that a dipolar modulation is completely equivalent to the Doppler effect on a map. We can thus make use of the peculiar velocity estimators derived in~\cite{Amendola:2010ty} (see eqs.~(B7), (B8) and (B15) therein) which give the cartesian components of a peculiar velocity from a map. Such estimators can be easily adapted to correspond also to the cartesian estimators of the modulation vector  by simply removing the aberration contribution to the estimator, leaving only the pure Doppler effect. In practice, this is achieved by using the following slightly modified version of (B8) of~\cite{Amendola:2010ty}:
\begin{equation}
\label{clmsT}
\begin{aligned}
    c_{\ell \,m}^{s\,+} & =  (-1)^{s} (-d)\sqrt{\frac{(\ell +1+m-s)(\ell +1-m+s)\left(\ell -s m+s^2\right)}{\left(4(\ell +1)^2-1\right)\left(1+s^2\right)(\ell +s m)}}\,, \\
    c_{\ell \,m}^{s\,-} & =  - d\,\sqrt{\frac{(\ell +m-s)(\ell -m+s)\left(\ell +s m+1-s^2\right)}{\left(4\ell ^2-1\right)\left(1+s^2\right)(\ell -s m+1)}}\,.
\end{aligned}
\end{equation}
Such an estimator has been derived only in the full-sky case limit and therefore it should be adapted to the partial sky case. One way would be to try to transform the pseudo-$a_{\ell m}$ into $a_{\ell m}$'s to use in~(B7) of~\cite{Amendola:2010ty}, but this would require to compute a MASTER-like tensor with 4 indices. We have checked however on masked boosted simulations that applying the estimator directly on the pseudo-$a_{\ell m}$'s and using the  pseudo-$C_\ell$'s everywhere in the coefficients of the estimator we obtain an unbiased estimate of the modulation. Therefore we simply used the pseudo-$C_\ell$'s everywhere, obtained by applying the MASTER matrix to the smooth best-fit Planck $C_\ell$'s.

Thus we obtain a modulation vector $\vec{A}_{mod}$ for each map and so we compare the results for the \emph{amplitude} $|\vec{A}_{mod}|$ of the modulation with our simulations. In an isotropic universe observed by a boosted observer 
the amplitude of the modulation should converge to a nonzero value due to the peculiar velocity. The presence of the anisotropic noise, moreover, biases the result at high $\ell$, so it is important to take it into account. We do this by summing the 100 Monte-Carlo simulated noise $a_{\ell m}$'s made available by the Planck collaboration and summing them (with the SMICA weights) to  our own simulated  temperature $a_{\ell m}$'s. Once more, we allow for two cases: one including Doppler and Aberration in the simulations, the other not.

\begin{figure}[t]
    $\!\!\!\!\!\!\!$\includegraphics[width=1.06\textwidth]{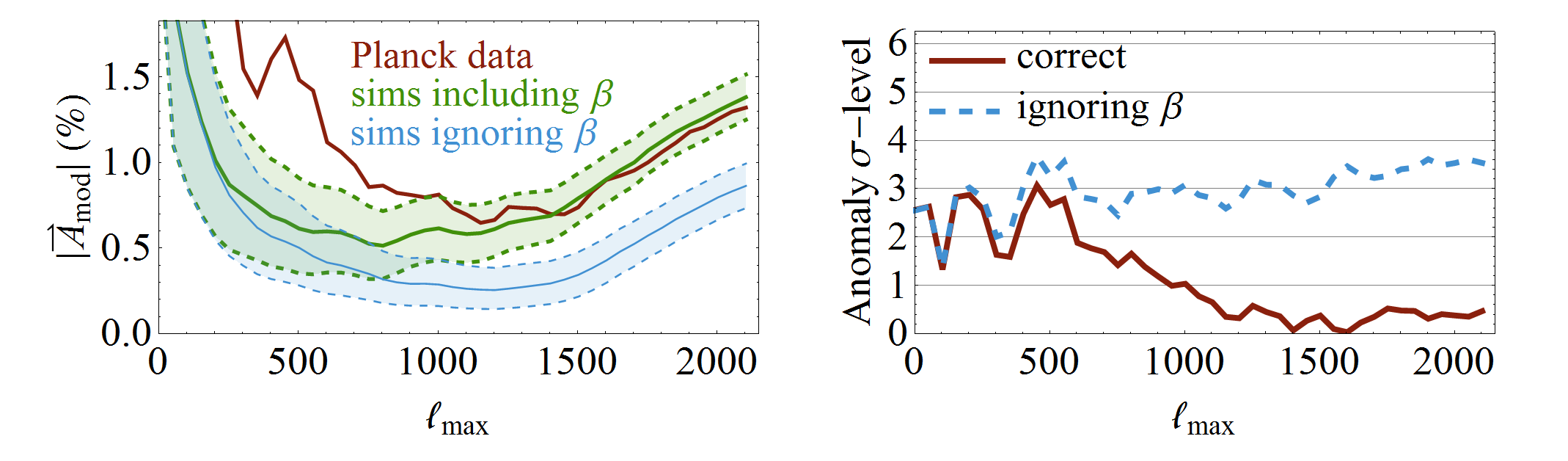}
    \caption{Similar to Figure~\ref{fig:amod-vs-ell} but without the {\emph a posteriori} selection of the direction of maximal anisotropy. The discrepancy here is typically below $2\sigma$ (but over $3\sigma$ if one neglects boost effects). The larger discrepancies are at larger scales ($\ell \in [2,600]$), which corresponds to the range of scales probed by WMAP. Note that here, contrary to the estimator in Figure~\ref{fig:amod-vs-ell} we do not subtract the noise asymmetry, which explains the increase of asymmetry on small scales, where the noise dominates. \label{fig:amod-vs-ell-betacart}}
\end{figure}

Figure~\ref{fig:amod-vs-ell-betacart} depicts the modulation amplitude obtained in this way in both cases. Each cartesian component of the modulation is Gaussianly distributed, and therefore the magnitude of the amplitude vector distribution is approximately given by a noncentral chi distribution. From this distribution we compute the $1\sigma$ bands as well as the significance levels of Figure~\ref{fig:amod-vs-ell-betacart}.\footnote{If we had access to all 1000 MC noise simulations from Planck, we could bypass the need to specify a noncentral chi distribution and rely on the interpolated histograms themselves.} As it is clear from the Figure, removing the \emph{a posteriori} selection of the maximal asymmetry direction leads to a diminished discrepancy between simulations and data, and we conclude that no significant modulation anomalies are present. The larger discrepancies are at large scales ($\ell \in [2,600]$), which corresponds to the range of scales probed by WMAP. These discrepancies however vanish when including smaller scales, unless one mistakenly ignored the boost. Note that some of the previous analyses have claimed~\cite{Ade:2013nlj,Flender:2013jja} that a {\it scale-dependent} modulation may be necessary to fit the data, in such a way that an effect is present at $\ell<600$ but it strongly decays at higher $\ell$. We stress here that such a conclusion is only weakly supported by our results, in the sense that there is no reason to split the data {\it a posteriori} into these two subsets and that our estimator is by construction scale-independent and therefore the whole range of data should be used. An analysis which includes scale dependence should be made with a different ansatz for the modulation, which hopefully should be physically motivated and with the least amount of extra parameters.

Note again that, as in the previous section, the boost is already non-negligible and alleviates the discrepancy by about half $\sigma$ even at $\ell<600$, which correspond to the WMAP scales of interest.

\section{Conclusions}  \label{sec:conclusions}

We have analyzed the CMB Temperature map released by Planck, the SMICA map, which is a linear weighted combination of different frequency maps and looked for differences in power spectra on opposite hemispheres  along some special directions: North South compared to the galactic plane, along the dipole direction and along the direction of \emph{maximal asymmetry} $(l,b)=(225^\circ,1^\circ)$ found by Planck~\cite{Ade:2013nlj} (which is also close to the WMAP maximal asymmetry direction~\cite{Eriksen:2003db,Eriksen:2007pc,Hansen:2008ym,Ade:2013nlj}).

We used a new technique to analyze anomalies by first building an antipodally symmetrized version of the U73 mask, which by construction cannot introduce a bias on hemispherical asymmetries, and then smoothed the resulting mask  on a $10'$ scale. We have thus reconstructed the real $C_\ell$'s using the MASTER approach and compared the real data with simulations. We properly added \emph{both} Doppler and aberration effects due to the peculiar motion of the observer using the methods developed in our previous work performed on simulations~\cite{Notari:2013iva} using them now on real data. We also added the anisotropic Planck noise. We find three main results, improving on previous analyses~\cite{Flender:2013jja} in several ways: with the symmetrized mask, the reconstruction of $C_\ell$'s from the $\widetilde{C}_\ell$'s, the inclusion of aberration and anisotropic noise and also by using a new estimator which finds the maximal asymmetry direction for each simulation.

First, all such effects are important, and after subtraction of the above effects we find that hemispherical anomalies are  less  significant than $1.5\sigma$ the full range of $\ell$, up to 2000.

Second, we find results on a scale-independent ``dipolar modulation'' $A_{\rm mod}$ of the power spectrum, by introducing a proper estimator, which differs by the one used in~\cite{Flender:2013jja} and takes into account of the boost and of anisotropic noise. Pre-selecting the {\it maximal asymmetry} direction (which was previously found by the Planck collaboration and it is consistent also with the maximal asymmetry direction of WMAP) we find  $ A_{\rm mod} = 0.0044\pm0.0014$  for the full range of $\,2\leq \ell \leq 2000$, which would constitute a $3.3\sigma$ anomaly when we compare to $A_{\rm mod}$ of simulations. Note that such value contrasts with the low-$\ell$ results~\cite{Eriksen:2003db,Eriksen:2007pc,Hansen:2008ym} including the Planck analysis~\cite{Ade:2013nlj}, which have both central value and error larger by an order of magnitude. However this $3.3\sigma$  result is an overestimate of the real statistical significance, when comparing with the modulation for simulations along a {\it generic} direction, while the selected direction for the Planck data was not randomly chosen. A better statistics which does not suffer from this bias is built by using an estimator which finds directly the {\it amplitude} of the modulation along the maximal asymmetry axis for each simulation. This is achieved by using three estimators which correspond to the cartesian components of a modulation vector $\vec{A}_{\rm mod}$. With this unbiased procedure we find that the real statistical significance in the range of $\,2\leq \ell \leq 2000$ is less than $1\sigma$, and it is only $2.4\sigma$,  in the WMAP-equivalent range  of $\,2\leq \ell \leq 600$. Note that the presence of a marginally significant modulation when considering $\ell <600$ scales and the absence of a modulation when including smaller scales may lead to the idea that it could be necessary to fit the data with a scale-dependent modulation. We stress here that such a conclusion is only weakly supported by our findings since we have no justification to split the data {\it a posteriori} into these two subsets for our  scale-independent estimators, and therefore the whole range of $\ell$ should be used. An analysis which includes scale dependence should be made with a different estimator, either because of a physically motivated reason or by showing that it improves the fit even at the expense of extra parameters.

Third, we stress that ignoring the effect of a boost would lead to a spurious significance of hemispherical anomalies at almost $3 \sigma$ in the full range of $\ell$'s
and an artificially enhanced  nonzero result for $A_{\rm mod}$, which would reach roughly the $3.5 \sigma$ level (and the 5$\sigma$ level without removing the above mentioned {\it a posteriori} bias).
As a last remark note also that a boost affects not only temperature maps but also polarization ones. Therefore future analysis of power asymmetries, for either temperature or polarization, should always take both Doppler and aberration effects into account.

\acknowledgments
Some of the results in this paper have been derived using the HEALPix package \cite{Gorski:2004by}. We thank Gabriela Antunes, Jacques Delabrouille, Jaume Garriga, Massimiliano Lattanzi, Michele Liguori, Paolo Natoli, Shaun Hotchkiss and Thiago Pereira for useful discussions and comments.  We thank Viviana Niro for discussions and suggestions (and checking the deboosted spectra). MQ is grateful to Brazilian research agencies CNPq and FAPERJ for support and to Universit\`a di Ferrara for hospitality. AN is grateful to  Universidade Federal do Rio de Janeiro (UFRJ) for hospitality. AN is supported by the grants EC FPA2010-20807-C02-02, AGAUR 2009-SGR-168.

\appendix
\section{Comparison of original and symmetrized masks} \label{sec:U73symm}

\begin{figure}[t]
\begin{minipage}[c]{\linewidth}
    \centering
    \includegraphics[width=\textwidth]{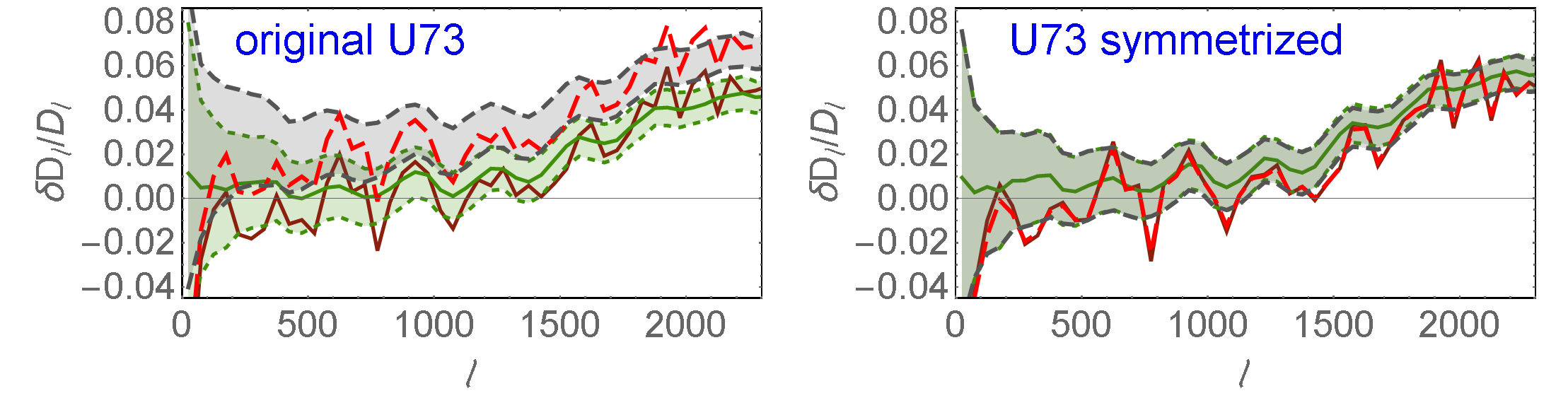}
\end{minipage}
    \caption{Corrections from symmetrizing the mask and applying the MASTER approach for the case of the hemispheres aligned with the galactic plane. The brown curve and green regions are the same as in Figure~\ref{fig:steps}. The dashed red curve and gray band stand for the same but without using the MASTER approach. We find that although the symmetrization of the mask removes the bias from the $\widetilde{C}_\ell$'s (i.e., prior to the MASTER correction), the difference becomes very small for the reconstructed $C_\ell$'s (i.e., after the MASTER correction). Computing the MASTER matrix is nevertheless time-consuming and its inverse become numerically unstable to compute unless one bins the multipoles. Therefore, using a symmetric mask might be more practical in some applications.}
    \label{fig:master-and-symm}
\end{figure}

In Figure~\ref{fig:master-and-symm} we illustrate the corrections introduced by symmetrizing the mask and employing the MASTER approach. As it can be seen, the non-symmetric U73 mask introduces a bias to the pseudo $\widetilde{D}_\ell$'s.  This bias is removed through the use of the MASTER matrix, in which case the end results are similar (but not identical) to the ones obtained using the symmetric U73symm mask. We conclude that the symmetrization of the mask introduces a small but not completely negligible correction. Since by construction it cancels any bias arising from leakage of mask power into the final result, we find it to be the safest approach.

It is interesting to note that for the complete masked sky, the use of a antipodally symmetric mask removes all first neighbors ($\ell,\,\ell+1$) correlations. This is however no longer true if we analyse individual hemispheres.

Another advantage of a symmetric mask arises in cases where multipole binning of the spectra is either undesirable or impractical, for instance when computing asymmetries in narrow $\ell$ ranges.
This is because the MASTER matrix become numerically unstable to invert unless one bins the multipoles. In those cases, using a symmetric mask as a safeguard could be even more important.

\section{Constructing a symmetrized masks} \label{sec:symmetrizing}
In order to construct a symmetric mask from a non-symmetric mask we work on pixel-space assuming the original mask is ordered according to HEALPix's RING ordering scheme. The simplest way to symmetrize the mask is to reorder the pixel indexing of the original mask to construct a parity-inverse mask and then multiply both masks.

This reordering can be achieved by setting each pixel angles $\{\theta, \phi\} \rightarrow \{\pi-\theta, \phi+\pi\}$. With these transformations ``polar'' pixels remain ``polar'' and likewise for ``equatorial'' pixels, so we can treat each region separately. Based on Eqs. (2)--(9) of~\cite{Gorski:2004by}, we define:
\begin{align}
    p_{\rm polar}^N(i,j) &\;=\; 2(i-1)i+j-1 \,,\\
    p_{\rm equat}(i,j) &\;=\; p_{\rm polar}^N(N_{\rm side}, 0) + 4 (i-N_{\rm side})N_{\rm side} + j \,,\\
    p_{\rm polar}^S(i,j) &\;=\; p_{\rm equat}(3N_{\rm side}+1, 0) - 2(i-5N_{\rm side})(i-1-3N_{\rm side}) +j \,,\\
    j_{\rm max}^N(i) &\;=\; 4i \,,\\
    j_{\rm max}^{\rm equat} &\;=\; 4 N_{\rm side}\, . \\
\end{align}
We then construct 2 lists, for the polar and equatorial strips:
\begin{equation}
\begin{aligned}
    &\texttt{For i = 1 to }N_{\rm side} - 1\\
    &\texttt{For j = 1 to } j_{\rm max}^N(i)\\
    &\mathrm{list1: }\;\bigg\{p_{\rm polar}^N(i,j),\,  p_{\rm polar}^S\Big(4 N_{\rm side} - i,\,
    1+ \mathrm{Mod}\Big(j+\frac{1}{2}j_{\rm max}^N(i)-1,j_{\rm max}^N(i)\Big)\Big)\bigg\}
\end{aligned}
\end{equation}
and
\begin{equation}
\begin{aligned}
    &\texttt{For i = $N_{\rm side}$ to }2N_{\rm side}\\
    &\texttt{For j = 1 to } j_{\rm max}^{\rm equat}\\
    &\mathrm{list2: }\;\bigg\{p_{\rm equat}(i,j),\,  p_{\rm equat}\Big(4 N_{\rm side} - i,\,
    1+ \mathrm{Mod}\Big(j+\frac{1}{2}j_{\rm max}^{\rm equat}-1,j_{\rm max}^{\rm equat}\Big)\Big)\bigg\}\,.
\end{aligned}
\end{equation}
Each list element is a pair of indices of antipodal pixels. The last $2N_{\rm side}$ terms of list2 are redundant, so we define a list3 by dropping these terms from list2. Finally, we construct one single list4 by the juxtaposition of list1 and list3. This final list represents the sought out indices pairings. By reordering the original mask pixels according to this list one gets an antipodal mask.

\section{Details of our HEALPix simulations} \label{sec:healpix}

In all our simulations we employ the following HEALPix parameters: \texttt{nside} = 2048; \texttt{nlmax} = 4000 (except for creating the masks, in which case we used \texttt{nlmax} = 5000); \texttt{iter\_order} = 1; \texttt{regression} = 0; \texttt{won} = 1; \texttt{iseed} = $-1$, \dots, $-500$. For Planck data and noise, we instead used $\texttt{iter\_order} = 3$. We also do not apply the pixel window functions. Our modified HEALPix allows the inclusion of boost effect in the \emph{synfast} routine. This is controlled through two new flags: ``\texttt{ab\_dopp}'' (0 for no boost, 1 for Doppler only, 2 for Doppler and aberration) and ``\texttt{beta}'' (the boost velocity). In order to employ a boost factor of 2.5, we apply 2 consecutive boosts, one with \texttt{ab\_dopp} = 2 and \texttt{beta} = 0.00123, and the other with \texttt{ab\_dopp} = 1 and \texttt{beta} = 0.001845. Finally, to deboost the SMICA map, we used the same procedure but inverting the sign of \texttt{beta}, taking care first to: (i) debeam the data, (ii) mask the sky (to avoid missing the foregrounds due to aberration) and (iii) remove the noise (which is added back at the end, avoiding a deboosting of the noise).


\bibliographystyle{JHEP}
\bibliography{asymmetry}

%

\end{document}